\documentclass[prl,twocolumn,english,showpacs]{revtex4-1}
\usepackage{amssymb}
\usepackage{color}
\usepackage{graphicx}
\usepackage{bbold}
\usepackage{bm}
\usepackage{amsmath}
\usepackage{amsfonts}
\usepackage{amssymb}
\usepackage{braket}

\newcommand{\cop}{c^{\phantom{\dagger}}}
\newcommand{\cdop}{c^{\dagger}}
\newcommand{\nop}{n}

\begin{document}

\title{Decay and revival of a transient trapped Fermi condensate}

\author{Timothy Harrison$^1$, Martin Link$^1$, Alexandra Behrle$^1$, Kuiyi Gao$^{1,3}$,  Andreas Kell$^1$, Johannes Kombe$^1$, Jean-S\'ebastien Bernier$^1$, Corinna Kollath$^1$, Michael K{\"ohl}$^1$}

\affiliation{$^1$Physikalisches Institut, University of Bonn, Wegelerstrasse 8, 53115 Bonn, Germany\\
$^3$Email: kuiyi@physik.uni-bonn.de}

\begin{abstract}
We study experimentally and theoretically the response of a two-component Fermi condensate in the strongly-interacting regime to a  quench of the interaction strength. The  quench is realized using a radio-frequency $\pi$-pulse to a third internal level with a different interaction strength. We find that the quench excites the monopole mode of the trap in the hydrodynamic regime and that an initial change of the condensate properties takes place on a time scale comparable or even larger than the quasi-particle relaxation time.  
\end{abstract}
\maketitle

Understanding the non-equilibrium behaviour of superfluids and superconductors is a very active field of
research. One aspect of particular interest relates to the capacity to switch, at will, superconductivity
on and off. Gaining such an ability would have clear applications in devices, but would also shed light
on pressing fundamental questions related to the dynamics of the pairing mechanism. Therefore, it is
not surprising that the physics associated with rapid parameter changes in superfluids and superconductors
has intrigued physicists for decades~\cite{Volkov1974,Yuzbashyan2015}. Famous examples include the strong drive of superconducting materials by short laser pulses~\cite{BasovHaule2011,Orenstein2012,ZhangAveritt2014,Giannetti2016}.
%
%
However, measurements and interpretations have remained challenging. One central difficulty to advance the fundamental understanding is to conduct quench experiments cleanly: in a real material it often is difficult to couple specifically to one degree of freedom, for example, to create excitations in the electronic sector without accidentally driving phonon modes simultaneously. This problem is further enhanced by the need for a very short perturbation necessary to reveal the (usually short-lived) genuine quantum dynamics. However, a short perturbation goes hand in hand with a broad excitation spectrum, which then couples to a wealth of excitations in the material. 

Ultracold atomic gases in atom traps offer a particularly clean realization of Fermi condensates. Additionally, the commonly employed technique of interaction control via Feshbach resonances allows for selective focusing on the ``electronic'' (corresponding to the fermionic gas) degree of freedom of the BCS problem in its purest form without invoking complications due to phonons. Therefore, ultracold Fermi condensates have been used to address the non-equilibrium physics of  BCS-type superfluids. For example, an interaction quench lead to the excitation of low-energy collective modes \cite{Riedl2008,Altmeyer2007,Tey2013} or the formation dynamics of a pair condensate \cite{Zwierlein2005b}. Yet, an interaction quench in a Fermi superfluid on time scales faster than the trap period and its ensuing quantum dynamics has not been studied.

 If we focus on the ``electronic'' sector, the dynamics of a quenched Fermi condensate can be categorized into two contributions: (1) the dynamics of the perturbed condensate and (2) the dynamics of quasi-particles, which are created by the fast parameter change. Initial theoretical work by Volkov and Kogan \cite{Volkov1974} considered the dynamics of a BCS superconductor after an infinitesimally small abrupt change of the superconducting gap energy $\Delta$. The subsequent coherent time evolution of the order parameter exhibits small-amplitude oscillations with a frequency of $2\Delta/\hbar$, which can be interpreted as the activation of the Higgs mode of the superconductor. In subsequent years, a large amount of work has been devoted to theoretically study and understand the effect of rapid quenches of superconducting states 
 \cite{BarankovSpivak2004, WarnerLeggett2005, SimonsBurnett2005, YuzbashyanAltshuler2006, BarankovLevitov2006, YuzbashyanDzero2006, PapenkortKuhn2007, DzeroAltshuler2007, Gurarie2009, DzeroAltshuler2009, Galitski2010, ScottStringari2012, YuzbashyanFoster2015}. One of the key findings has been that in a collisionless superconducting regime the behaviour at short times can be described  by the (integrable) BCS model and that in the long-time limit the dynamics reduces to even simpler effective Hamiltonians with only few collective (spin) degrees of freedom, even for strong excitations \cite{Barankov2004,Yuzbashyan2015,Yuzbashyan2006b,Barankov2006,Yuzbashyan2006}. This has lead to the remarkable observation that, even though the physical situation is a genuine quantum many-body problem, the time evolution can exhibit large anharmonic oscillations of the order parameter, which are reminiscent of the physics of collapse and revival known from few-particle dynamics \cite{Milburn1986}. More precisely, the rapid perturbation of a three-dimensional homogeneous system of interacting fermions in the BCS regime has been predicted to exhibit distinct dynamical phases \cite{Yuzbashyan2015}, see Figure 1b. In ``Phase I'' the superfluid order parameter disappears rapidly \cite{Yuzbashyan2006b,Barankov2006}, ``Phase II'' is characterized by damped oscillation \cite{Volkov1974}, and ``Phase III'' is signaled by a persistent oscillation with a constant amplitude \cite{Barankov2004,Barankov2006,Yuzbashyan2006}.  Extensions of this work to harmonically trapped gases have generally confirmed the qualitative picture \cite{Hannibal2018}, however, with modifications in detail \cite{Scott2012}. The quench in a BCS state were also investigated for interaction changes using a linear ramp with a finite ramp time \cite{SentefKollath2016,Kombe2019, Mazza2017}. Some of the features discovered for the sudden ramp were found to survive in a mean-field treamtent of a finite but still rapid ramp time \cite{SentefKollath2016,Kombe2019}.

However, real experiments with Fermi condensates in ultracold gases are not in a collisionless regime but typically close to the unitary regime with, say, $|1/(k_Fa)| \lesssim 1$, and correspondingly large quasiparticle scattering rates. Here, $k_F$ denotes the Fermi wave vector and $a$ the $s$--wave scattering length. How much of the physics of the integrable theory is applicable in this regime is still unclear as the potential observability of any coherent dynamics of the condensate depends critically on the relaxation mechanisms and lifetimes of collective and quasiparticle excitations.
%
%
Only precisely at unitarity, $1/(k_Fa)=0$, theoretical predictions have been made for a so-called holographic superconductor using AdS/CFT correspondence and found oscillating or steady-state final states depending on the strength of the quench \cite{Bhaseen2013}.

In this manuscript, we study the non-equilibrium physics of a fermionic superfluid made from ultracold atomic gases near the unitary regime. We study the quench from a strongly-interacting initial state to a weakly-interacting final state realized by the application of a radio-frequency (rf) $\pi$-pulse. In previous experiments rf-pulses were mainly used in order to investigate the equilibrium phases of the Fermi gas \cite{GreinerJin2005,Chin2004,Shin2007,Stewart2008,Feld2011} and only recently, the excitation of a Higgs mode \cite{Behrle2018} has been probed.
We reveal a dynamics of collapse and revival of the condensate together with the appearance of collective modes. The time scales observed for the collapse and revival of the condensate are very long and rather more comparable to the trap period than to the intrinsic time scales of the superfluid. By numerical modelling of the full dynamics in a one-dimensional system including the third state and final state interactions, we confirm that this effect results from the interplay between density inhomogeneity and pairing dynamics. 

Our measurements are conducted in an ultracold quantum gas of $ 10^6$ $^6$Li atoms prepared in a balanced mixture of the two lowest hyperfine states $\ket{1}$ and $\ket{2}$ of the electronic ground state $^2S_{1/2}$ \cite{Behrle2018}. The gas is trapped in a harmonic potential with frequencies of $(\omega_x,\omega_y,\omega_z)=2\pi \times (110,151,234) \,$Hz and is subjected to a homogeneous magnetic field in the range of $880-1000$\,G  in order to tune the $s$--wave scattering length $a$ near the Feshbach resonance located at 834\,G. This results in an adjustment of the initial interaction parameter of the gas in the range of $-0.7\lesssim 1/(k_Fa_i)\lesssim -0.1$, i.e. on the BCS side of the BCS-BEC crossover. The Fermi energy in the center of the gas is $E_F\simeq h\times (29\pm 3)$\,kHz at each of the considered interaction strengths and sets the Fermi wave vector $k_F=\sqrt{8 \pi^2 m E_F/h^2}$, where $m$ denotes the mass of the atom and $h$ is Planck's constant. 

\begin{figure}
 \includegraphics[width=\columnwidth,clip=true]{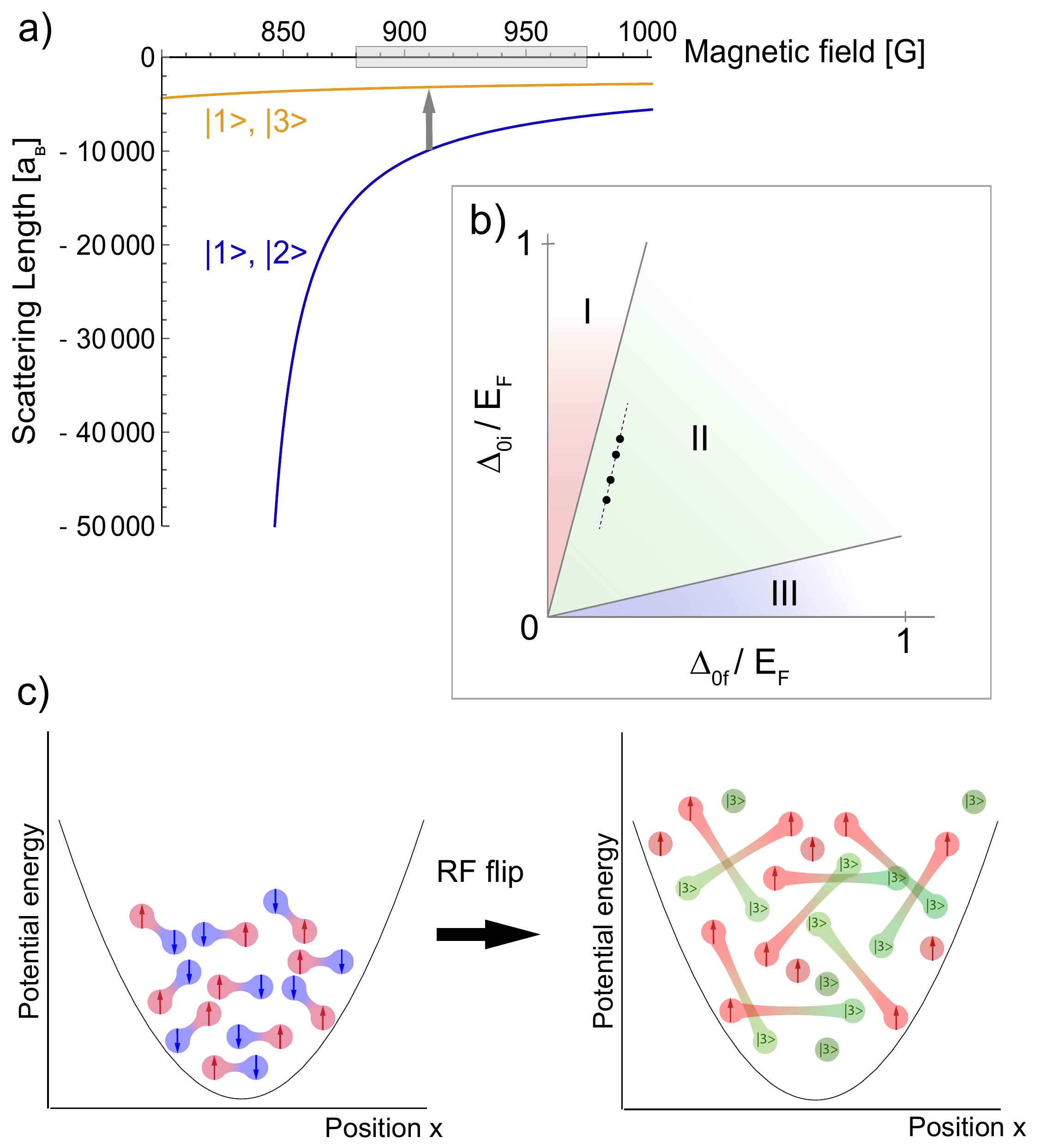}
 \caption{ Schematic of the quench experiment. a) The Feshbach resonances of the $\ket{1}, \ket{2}$ and $\ket{1},\ket{3}$ mixtures. The grey-shaded region shows the range of considered interaction strengths. The gas is switched from the strongly-interacting $12$-mixture of lithium into the weakly-interacting $13$-mixture by a radio-frequency pulse. This induces a rapid change of the s-wave scattering length $a$. b) Sketch of the mean-field quench phase diagram of the collisionless superfluid \cite{Yuzbashyan2015} together with indications  of our initial and final equilibrium gap values $\Delta_{0i}$ and $\Delta_{0f}$, respectively.  c) Sketch of the equilibrium  distributions of the trapped superfluid for the initial and final interaction strength including the breaking of pairs. }
\label{fig1}
\end{figure}

Performing an instantaneous quench of the interaction is very challenging since, typically, magnetic fields of several ten or hundred Gauss strength would need to be varied in  a few microseconds, which usually is hindered by practical constraints such as eddy currents. In order to obtain a relatively quick change of the interaction strength, we therefore follow a different route: we perform a  quench from the strongly-interacting superfluid in the $\ket{1}$ and $\ket{2}$ states to the relatively weakly-interacting mixture of the $\ket{1}$ and $\ket{3}$ states (see Fig. \ref{fig1}a) by a  radio-frequency $\pi-$pulse  with a duration $\tau=28(2)\,\mu \text{s}=4.8\times \hbar/E_F$ and a  transfer efficiency of  $>97\%$ at 910\,G. A quench of this duration is slow as compared to the internal time scales $\sim \hbar/E_F$ but fast as compared to the time scales of the harmonic trap $\sim 2\pi/\omega$. For smaller values of the magnetic field than 880\,G, i.e., closer to unitarity, we observe atom losses and we refrain from performing quench experiments in this domain. Qualitatively, the situation at the quench time can be considered as follows: The initial state at interaction strength $1/(k_Fa_i)$ has a superconducting gap of $\Delta_i$ and a chemical potential $\mu$ which govern the density profile in the trap. After the quench to a weaker interaction strength $1/(k_Fa_f)$ the gap is reduced to a nominal gap value of $\Delta_f$ and the chemical potential increases, see Figures 1b and c. Both effects trigger internal dynamics and an adjustment of the density distribution, and we monitor the subsequent time evolution. To this end, we perform a magnetic field sweep (``rapid ramp'') from the BCS side of the Feshbach resonance to the BEC side and convert Cooper pairs into tightly-bound molecules \cite{Regal2004a}. We extract the data from absorption images taken after a ballistic expansion of 15\,ms. The magnetic bias field  gives rise to a weak harmonic confinement during the ballistic expansion and we have chosen the expansion time corresponding to a quarter oscillation period in this potential and hence our data reflects the momentum distribution. We fit the data with a bimodal distribution comprising of two Gaussians in order to assess the dynamics of the condensed part and the uncondensed part individually. For more detail, see Supplemental Material

\begin{figure}
 \includegraphics[width=\columnwidth,clip=true]{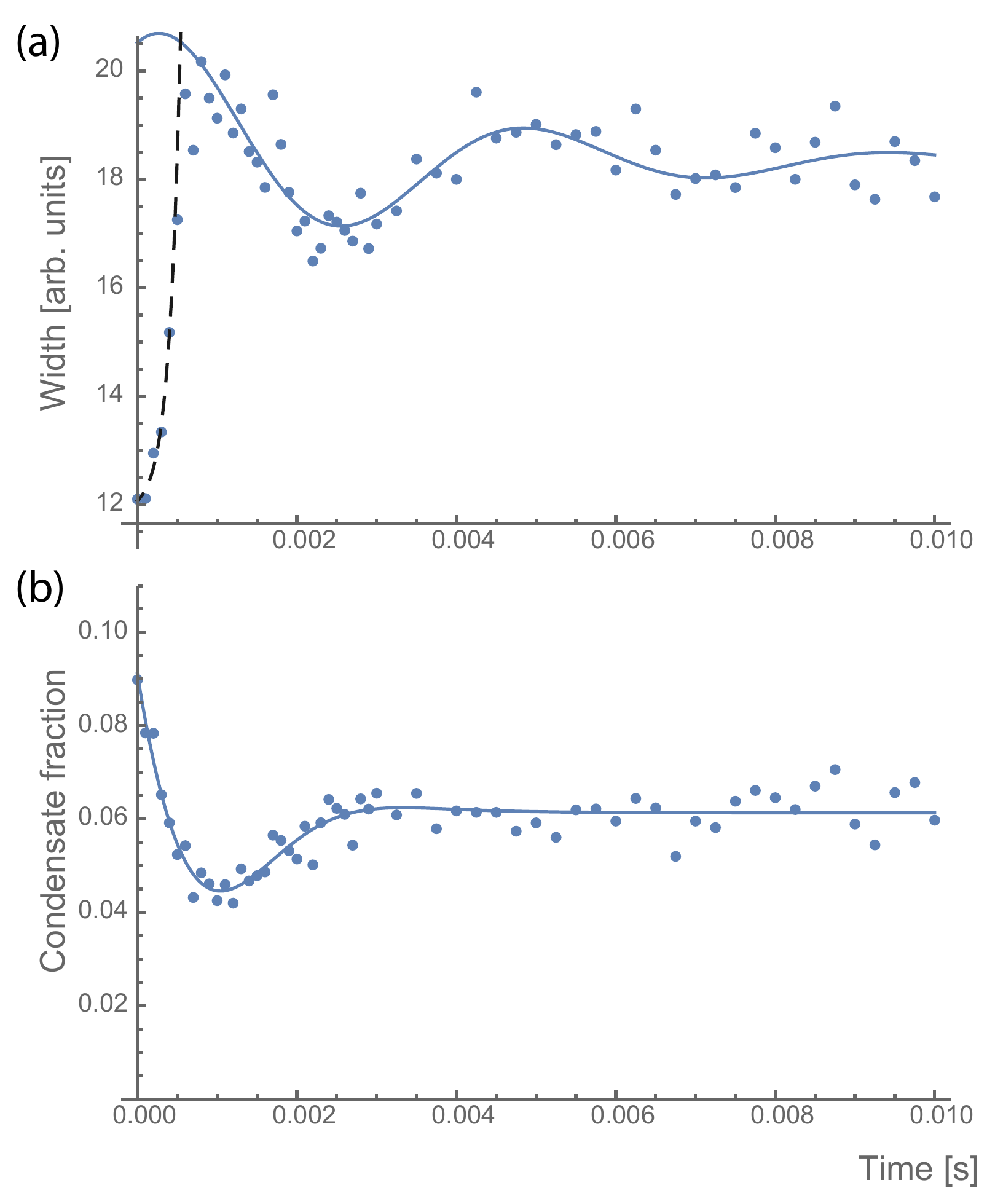}
 \caption{Time evolution after the quench  from $1/(k_Fa_0)=-0.74$ to $1/(k_Fa_1)=-1.31$. (a) Width of the condensate peak (blue points). The dashed line shows the exponential growth fit in the early stages of the time evolution and the solid line the damped oscillation fit to the longer times. (b) Condensate fraction (blue data points) together with the fit described in the text (solid line).}
\label{fig2}
\end{figure}

In Figure \ref{fig2} we show, exemplary, the recorded condensate dynamics as a function of the wait time after a quench from $1/(k_Fa_0)=-0.74$ to $1/(k_Fa_1)=-1.31$.
We consider the evolution of both the width of the condensate peak and the condensate fraction. Qualitatively, we find very similar behaviour for all quenches studied. 

First, we observe that the interaction quench induces a fast increase of the width of the condensate momentum distribution which  triggers  a  collective mode displayed as periodic oscillations in the width of the momentum distribution for both the condensate and thermal parts. These oscillations are expected in the harmonic trap  since the  interaction quench is fast as compared to the  time scale of the trap and produces a perturbation proportional to the initial density distribution. Hence the perturbation has the perfect symmetry to induce a monopole mode. We fit the dynamics  beyond the fast initial rise with an exponentially decaying harmonic oscillation in order to extract the frequency. The measured frequencies for the different final interaction strengths are shown in Figure 3 and we do not see a significant variation across the covered interaction range. We compare the measured mode frequency with a hydrodynamic model \cite{Kagan1996,Castin1996}, which we solve numerically for our trap geometry. The hydrodynamic model does not include details of the interaction but only the equation of state of the unitary gas and hence the oscillation frequency we derive is independent of $1/(k_Fa)$. We find very good agreement of the measured frequency  with the theoretical prediction of the monopole mode, verifying that our data are in the hydrodynamic and not in the collisionless BCS regime. The width of the distribution of the thermal pairs also exhibits  oscillations of the same frequency as the condensate part, which is expected in the hydrodynamic regime. Moreover, we have experimentally varied the confinement of the gas and find the corresponding scaling of the monopole frequency.

Second, we study the fast initial rise of the width of the condensate momentum distribution, the corresponding signal of which is absent in the thermal cloud. This increase occurs on a time scale much faster than the period of the monopole oscillation. In order to extract the timescale of this dynamics, we fit the first 400\,$\mu$s of the time evolution with an exponentially increasing function $w_0+w_1 \exp(t/\tau_0)$. We plot the extracted time scales $\tau_0$ in Figure 3. The detected time scale is very much comparable to the quasiparticle relaxation time $\hbar E_F/\Delta_f^2$, which we estimate using the mean-field equilibrium for $\Delta_f$.  This suggests that, after the interaction quench, excess quasiparticles redistribute. This is the fastest dynamics we observe; it occurs on a timescale approximately 30 times shorter than the collective mode.

Third, we observe that the condensate fraction shows a pronounced time dependence. Initially, the condensate fraction decays sharply and goes through a minimum value before reviving to reach a steady state on a time scale of $\sim 500\times \hbar/E_F$. We fit the data of the condensate fraction with an exponential decay for the initial period and a stretched exponential for the revival $A \exp[-t/\tau_1]+B(1-\exp[-(t/\tau_2)^\gamma])$ in order to extract the decay time $\tau_1$ and the rise time $\tau_2$ together with the initial and final values of the condensate fraction $A$ and $B$, respectively. The stretching exponent has been fitted to $\gamma=2.1\pm 0.1$ across all data sets. The decay time $\tau_1$ depends weakly on the interaction strength and the decay time is approximately a factor of two shorter than the time of the recovery, $\tau_2$, and a factor two to three longer than the quasiparticle relaxation time.

The observed revival of the condensate is unexpected according to prior theoretical analysis of a slow quench of a collisionless superfluid~\cite{Sentef2016, Kombe2019}, which -- in contrast to rapid quenches \cite{Volkov1974,Scott2012} -- showed the absence of oscillations of the order parameter. Additionally, previous measurements of collective modes -- including the monopole mode -- have not reported simultaneous oscillations on the condensate fraction \cite{Altmeyer2007,Vogt2012,Baur2013,Tey2013,Holten2018,Peppler2018}. 
The time evolution of the superfluid is surprisingly slow, and the time scale of the revival is comparable to the trapping frequency rather than the quasiparticle relaxation time. Moreover, we find that the condensate spread nearly doubles during the period over which the condensate fraction diminishes. In comparison, the momentum spread of thermal background also increases in width but only by less than ten percent.  

\begin{figure}
	\includegraphics[width=\columnwidth,clip=true]{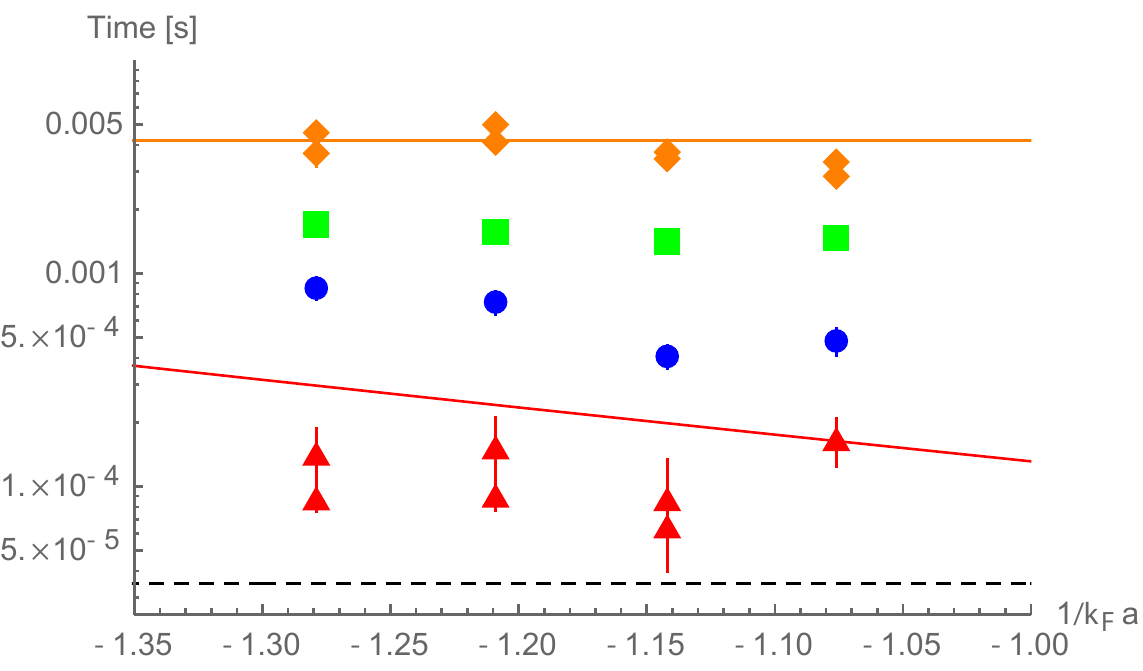}
	\caption{Time scales of the condensate dynamics. Orange diamonds: measured monopole oscillation period, solid orange line: prediction of the hydrodynamic model, green squares: $\tau_2$ (condensate fraction revival time), blue circles: $\tau_1$ (condensate fraction decay time), red triangles: $\tau_0$ (condensate width initial rise), red line: quasiparticle relaxation time, dashed black line: duration of the quench. For each quantity we show two symbols corresponding to different spatial directions.  Error bars denote the fit error; where not visible, they are smaller than the data point.}
	\label{fig3}
\end{figure}

\begin{figure}
\includegraphics[width=\columnwidth,clip=true]{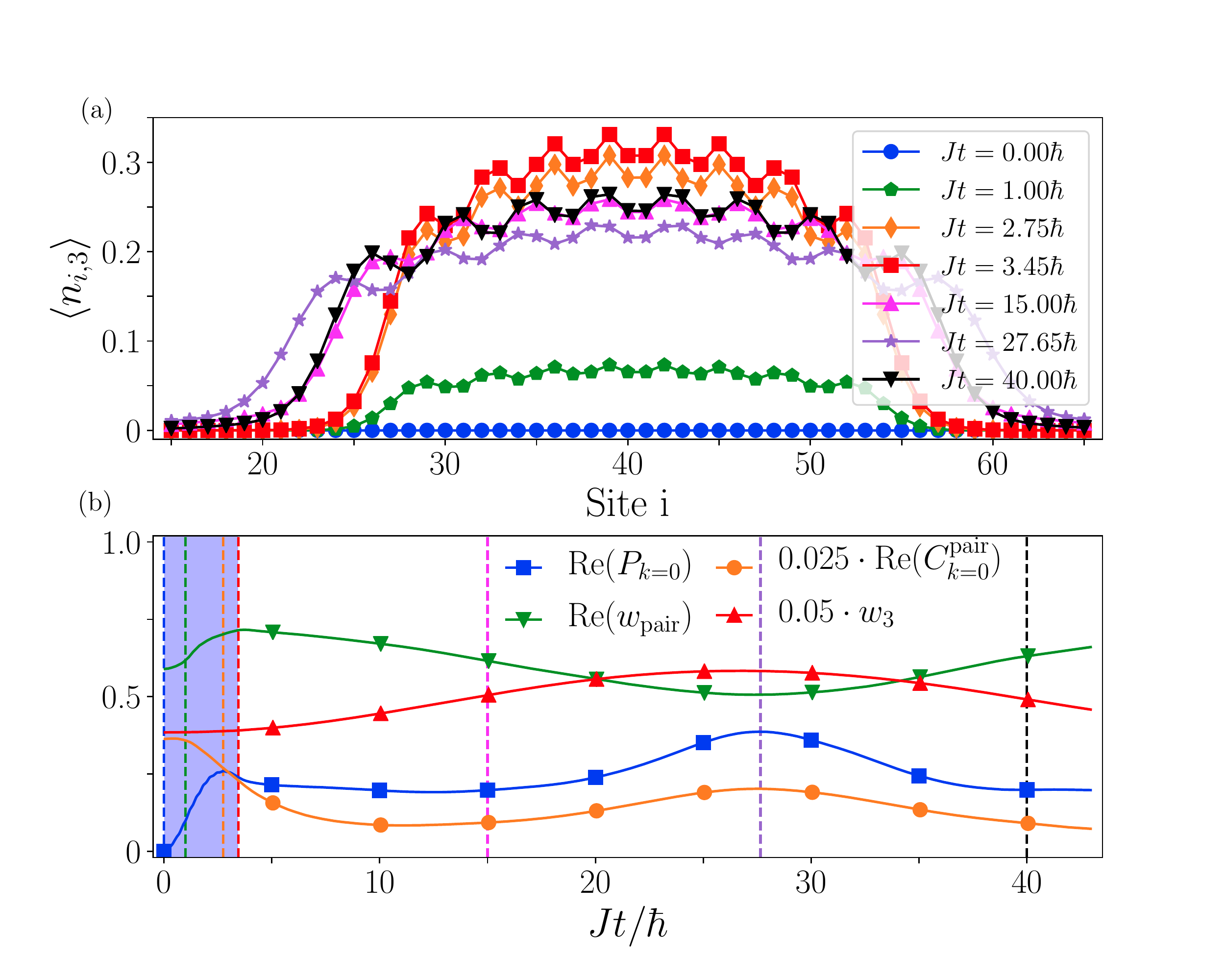}
\caption{(a) Snapshots of the density distribution of  state $\ket{3}$ in the trap starting from the empty state at time $t=0$.
  (b) Time-evolution of different observables: $w_3$ is the width of the density distribution of state $\ket{3}$,
  $\textrm{Re} (P_{k=0})$ is the coherence between pairs formed of atoms in states $\ket{1}$ and $\ket{3}$,
  $w_{\textrm{pair}}$ is the width of the pair correlation distribution in momentum space, $C^{\textrm{pair}}_{k=0}$
  is the pair coherence rescaled by the pair density (see method section for the definitions of the observables). The shaded area marks the duration of the rf-pulse. The dashed vertical lines mark the times shown in panel (a).
  We chose a lattice of $N=80$ sites and $N_{1}=N_{2}=8$ particles in level one and two in the initial state. The interaction parameters
  are chosen as $U_{12}=-6J$, $U_{13}=-2J$, $U_{23}=-2J$. The energy offset to the third level is $V_{3} = 50J$, the rf-driving frequency $\hbar\omega_{\text{rf}}=52.9J$,  amplitude $\hbar\Omega_{23} = 1.0J$, and pulse duration $Jt_{drive}=3.45\hbar$. A trapping potential $V_{\textrm{trap}} = 0.001J$ is present
  for all levels. We verified that a good convergence of our numerical results is achieved.}
  \label{fig:dmrg}
\end{figure}

In order to elucidate the dynamics and countercheck our interpretation, we perform numerical simulations of a Fermi gas modeled as an attractive Hubbard model with three internal species. We use a low filling in order to minimize lattice effects and focus on one spatial dimension in order to make an exact treatment by the time-dependent density matrix renormalization group method possible \cite{DaleyVidal2004,WhiteFeiguin2004, Schollwoeck2011}. Here we go beyond most previous treatments by taking into account (1) the interactions in the initial {\it and} final states of the three-state problem, and (2) the explicit time-dependence of the state change due to the radio-frequency $\pi$--pulse. 
The results of the time-evolution, including the time during which the rf-drive is applied, are shown in Figure \ref{fig:dmrg}. The shaded region marks the duration of the rf-drive ending at $tJ = 3.45\hbar$.

The initially empty level $\ket{3}$ is populated under the effect of the rf-drive resulting in a
strong increase of the density in level $\ket{3}$ (Figure~\ref{fig:dmrg}a). After the rf-drive is turned off
at $t J = 3.45\hbar$, the number of atoms in state $\ket{3}$ remains constant. Nevertheless,
atoms continue to redistribute within the trap as the density profile broadens up to $t J =27.65\hbar$ and
then contracts afterwards. 
%
%
This signals the excitation of the collective monopole mode by the rf-induced interaction quench. During the application of the rf-drive, as the third state becomes occupied, the number of pairs
formed between the states $\ket{1}$ and $\ket{3}$, as well as the pair coherence, $\textrm{Re} (P_{k=0})$,
and the second moment, $w^2_{\textrm{pair}}$, of the pair correlation momentum distribution display a strong
initial increase (Figure~\ref{fig:dmrg}b). One should note that these last two quantities are related to the
experimental condensate fraction and to the width of the condensate, respectively. However, the pair coherence
begins to decrease before the end of the rf-drive. This decrease is due to the loss of coherence over
longer distances which is no longer compensated by the increase in the number of atoms in level $\ket{3}$.
To be more precise, we consider the long distance pair distribution rescaled by the pair density, $C^{\textrm{pair}}_{k=0}$. This quantity, uncluttered with the pair density dynamics, presents a strong decrease at short times
highlighting the decay of the pair coherence. At later times, the pair coherence oscillates with a
maximum at precisely the time when the density distribution becomes broadest, allowing for longer range coherence.
Thus, these long pair coherences measured by the zero-momentum peak
are influenced by the monopole oscillations in the density, and therefore also show the oscillations with
the monopole frequency. The oscillations of the width of the pair momentum peak and its amplitude are not in phase.
The dynamical effects uncovered within our theoretical model, are in good agreement with the experimental findings
presented earlier and support strongly the interpretation provided above. 

In conclusion, we have studied the dynamics of a Fermi condensate near the BCS/BEC crossover subject to an interaction quench. We find that the monopole dynamics of the trap is excited in the hydrodynamic regime by the interaction quench. Additionally, the condensate dynamics shows an initial decay on a time scale comparable or larger than the quasi-particle relaxation time. After this initial decay, the condensate fraction subsequently stabilizes to a steady-state. The experimental findings are confirmed by theoretical results for a one-dimensional system which fully models the dynamics of the three internal levels taking final state interactions into account.

The work has been supported by the Alexander-von-Humboldt Stiftung, ERC (grants 616082 [K.G., M.K.] and 648166 [C.K, J.K., J.-S. B.]), Studienstiftung des deutschen Volkes [T.H.], Deutsche Forschungsgemeinschaft (DFG, German Research Foundation) under project number 277625399 - TRR 185 (B4)  [A.B., M.L.,M.K.,C.K.] and project number 277146847 - CRC 1238 (C05) [J.-S. B., C.K.] and  under Germany's Excellence Strategy – Cluster of Excellence Matter and Light for Quantum Computing (ML4Q) EXC 2004/1 – 390534769 [M.K., C.K.].



%

\section{Supplemental Material}

\textbf{Preparation.} In our experiment, the strongly interacting Fermi gas in an equal mixture of the two lowest hyperfine states $\ket{1}$ and $\ket{2}$ of $^{6}$Li atoms, is initially evaporatively cooled in a crossed-beam optical dipole trap. The trap frequencies during the interaction quench and the evolution stage are $2\pi \cdot (110, 151, 235)$ Hz. The sample contains $\sim 1 \times 10^6$ atoms for each spin component, and the temperature is $T/T_F=0.07\pm0.02$. The Fermi energy is $h \cdot 29(3) $ kHz. The magnetic field is set to 795 G for an efficient evaporation and slowly ramps in 200 ms to a final field with subsequent 200 ms hold time for equilibration.  The final value of the magnetic field ranges from 834 to 1000 G, which sets the amplitude of the change of the interaction strength.

\begin{table}[h]
	\centering
	\caption{\label{tab:table1} Measured transfer efficiency of the rf-flip from $\ket{2}$ to $\ket{3}$ state.}
	
	\begin{tabular}{|c|c|}
		\hline Final magnetic field (G) & Transfer efficiency (\%) \\
		\hline 880 & 88.0 \\
		\hline 895 & 90.7 \\
		\hline 910 & 97.2 \\
		\hline 1000 & 98.4 \\
		\hline
	\end{tabular}
	
\end{table}


\textbf{Detection.} 
We detect the evolution of the transient trapped Fermi condensate by rapidly extinguishing the optical dipole trap  and apply a rapid ramp from the final magnetic field to the zero-crossing of the scattering length at 569 G of the 13 mixture. Therefore, during expansion of the Fermi gas, the distortion of the momentum distribution due to the interaction effect is minimized. The density distribution after a quarter period of the weak Feshbach curvature trap is a mapping of the initial momentum distribution in the hybrid trap.

\textbf{Bimodal fitting.} To quantitatively analyse the evolution of the transient trapped Fermi condensate, a bimodal fitting of the momentum distribution after the expansion is implemented.  In order to fit the thermal fraction of the sample, initially, the centre of the distribution is cut out. The boundary for this cut comes from a simple bimodal fit, but the size of the masked out centre region can be varied. Then, a two dimensional Gaussian fit is implemented to the remaining data. This makes sure, that only the wings and therefore the thermal part of the cloud is fitted. For the next step, the thermal background is subtracted, so that only the condensed part remains. To perform this, another two dimensional Gaussian fit is applied. From the fitted parameters of the Gaussian distributions, some important quantities, such as the condensate fraction $N_0/N$ as and momentum distributions for both condensate and thermal parts, can be calculated.

\textbf{Theoretical treatment}
We consider a Fermi-Hubbard model at zero temperature with three internal species which are attractively interacting in order to model
the experimental setup. The Hamiltonian is given by
\begin{eqnarray}
H_0 =&& - J \sum_{\langle i,j \rangle, \sigma} \cdop_{i,\sigma}\cop_{j,\sigma}
+ \sum_{\sigma,i=1}^{L} V_{\text{trap}} \Big( i - \frac{L+1}{2} \Big)^2 \nop_{i,\sigma} \nonumber \\
&& + \sum_{i=1}^{L} \left\{ \sum_{\sigma < \tau} U_{\sigma,\tau} \nop_{i,\sigma}\nop_{i,\tau} + V_{3}\nop_{i,3} \right\},
\nonumber 
\end{eqnarray}
%
where $\hat{c}^{(\dagger)}_{i,\sigma}$ are the fermionic annihilation (creation) operators for states $\sigma = \{1,2,3\}$ on site i, 
$\nop_{i,\sigma}$ the corresponding number operator, and $\langle i,j \rangle$ denotes a sum over nearest neighbours in the 
one-dimensional lattice. $J$ denotes the hopping amplitude, $U_{\sigma,\tau} < 0$ the attractive on-site interaction, 
and $L$ the number of lattice sites. Typically, the energetic splitting $V_{3}$ between the state $\ket{2}$ and $\ket{3}$
is usually much larger than the kinetic and interaction energy scales, i.e. $V_{3} \gg J, U$.

Initially, at time $t=0$ the fermions are prepared in an equal mixture of state $\sigma=1,2$.
As we model a continuous Fermi gas in the absence of a lattice, for the results shown in Fig.~4 of the main text, we chose a very low density of fermions for which, typically, a
continuum approximation would be valid.
%
%
More specifically, we chose $L=80$, and $N_1=N_2=8$ and a trap of strength $V_{\text{trap}} = 0.001J$ confining
the fermionic gas to the central $\sim 30$ sites in the initial state. Then a pulse is applied using a rf-field. 
This rf-coupling induces mainly transitions between the internal states $\ket{2}$ and $\ket{3}$ of the atoms and
can be modelled by the term 
\begin{eqnarray}
H'(t) &=& \hbar \Omega_{23} \cos(\omega_{\text{rf}}t) \sum_{i=1}^{L} (\cdop_{i,3} \cop_{i,2} + \text{h.c.}), \nonumber
\end{eqnarray}
where $\Omega_{23}$ is the Rabi frequency of the transition (related to the dipole matrix element) and $\omega_{\text{rf}}$
the frequency of the rf-field.

\begin{figure}
	\includegraphics[width=\columnwidth,clip=true]{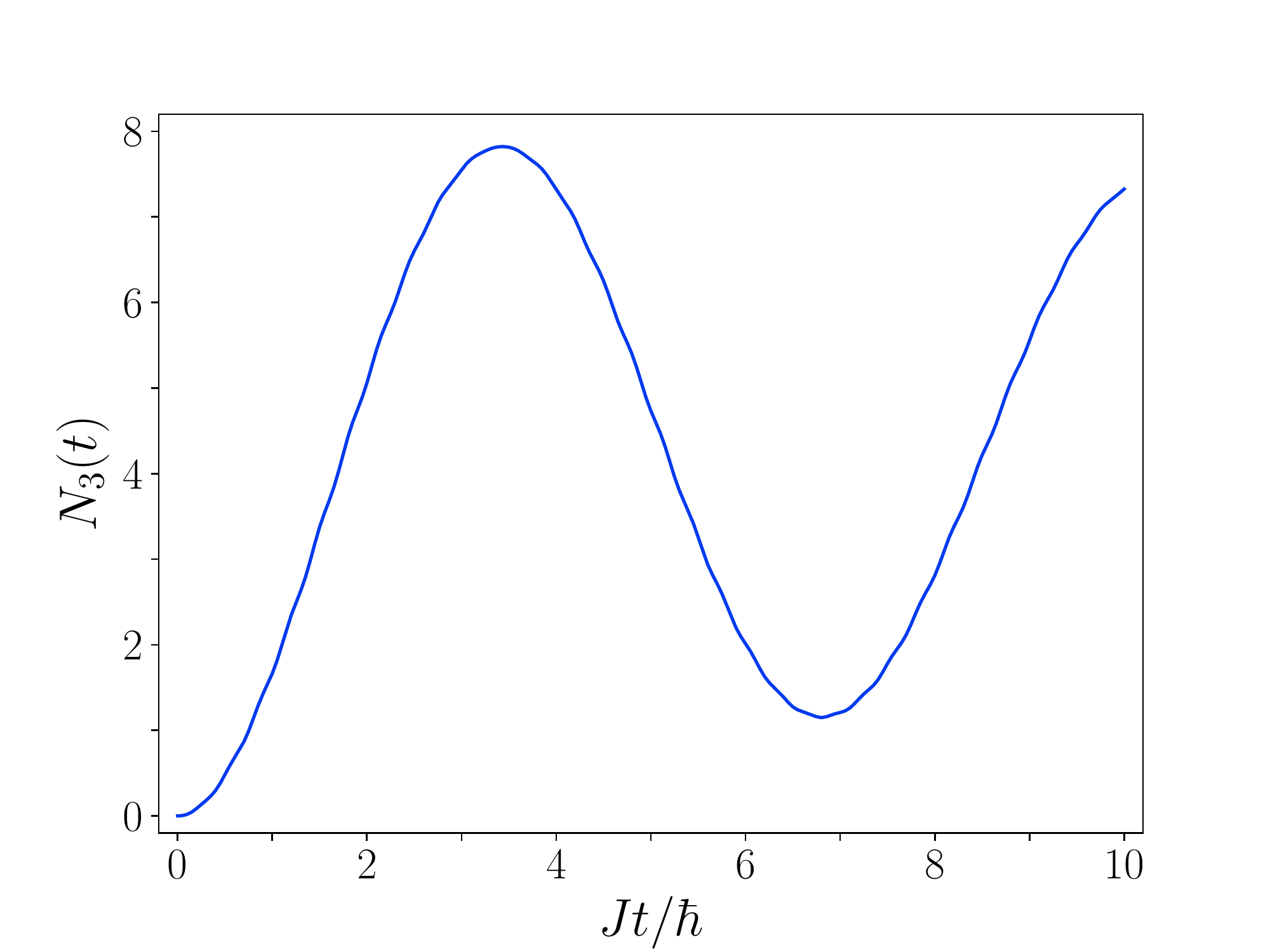}
	\caption{Time evolution of the total transfer $N_3$ to state $3$. The rf-pulse used in the simulation presented in Fig. 4 of the main text is stopped at the time $Jt = 3.45\hbar$ achieving maximal transfer.}
	\label{fig:dmrgN3}
\end{figure}

In order to calibrate the required parameters for a $\pi$--pulse in this interacting fermionic gas, we monitor the time-evolution
of the upper level population, $N_{3}(t)$, for different driving frequencies $\omega_{\text{rf}}$. The time-evolution of $N_{3}(t)$
exhibits distinct regimes reaching from Rabi-like oscillations as expected for a two-level system to an almost linear rise
as expected for the coupling to a continuum. However, at short times a maximal transfer can be identified around
$\hbar \omega_{\text{rf}} \sim 52.9J$, where the time-evolution resembles Rabi-oscillations (see Fig.~\ref{fig:dmrgN3}).

Here an almost perfect $\pi$--pulse with a transfer of $98\%$  (maximal value of $N_{3} \sim 7.8$) is reached for a pulse duration of $Jt = 3.45\hbar$.
One should note that the 'resonance' position is shifted from the non-interacting two-level system, where it would
occur at $\hbar \omega_{\text{rf}} = 50J$.

In order to characterize the evolution taking place during and after the effective interaction quench due to
the $\pi$--pulse, we define the following observables.
We consider the second moment of the density distribution of the third level which provides
a measure of the width of the distribution and is defined as
\begin{equation}
w^2_{3}(t) =  \frac{1}{N_{3}(t)}\sum_{i} (i - i_{c})^{2} \langle \nop_{i,3}\rangle (t), \nonumber
\end{equation}
where $i_{c} = (L+1)/2$ is the center of the lattice.

The pair coherence provides valuable information about the evolution of the system. We therefore monitor how
fast it builds up between the states $\ket{1}$ and $\ket{3}$ considering the $k=0$ amplitude of the
pair correlation. We define the pair correlation in momentum space as
\begin{equation}
P_{k}(t) = \frac{1}{L} \sum_{i,j} e^{ik(r_{i} - r_{j})} 
\langle \hat{\Delta}^{\dagger}_{i} \hat{\Delta}^{\phantom{\dagger}}_{j}  \rangle,
\nonumber
\end{equation}
the pair
annihilation operator $\hat{\Delta}^{\phantom{\dagger}}_{i} = \cop_{i,1}\cop_{i,3}$ at site $i$ with $k = \frac{2n\pi}{L}$ and $n = \{-L/2+1, \ldots, L/2\}$.

In order to measure also the width of the pair coherence, we define
\begin{equation}
w^2_\text{pair} = \frac{1}{P(t)}\sum_{k} k^{2} P_{k}(t) = \langle k^{2} \rangle, \nonumber
\end{equation}
where $P(t) = \sum_{k} P_{k}$ is the total number of pairs between $\ket{1}$ and $\ket{3}$.

As in the system under study the density dynamics induced by the trap plays a crucial role, we consider
a scaled pair coherence where each local pair correlator is divided by the time-dependent pair density at the corresponding spatial location such that
\begin{equation}
C^{\text{pair}}_{k=0} = \frac{1}{L_{\text{eff}}} \sum_{20 < i,j < 61 }
\frac{ \langle \hat{\Delta}^{\dagger}_{i} \hat{\Delta}^{\phantom{\dagger}}_{j} \rangle}{ \sqrt{\langle \nop_{i,1}\nop_{i,3}\rangle \langle \nop_{j,1}\nop_{j,3}}\rangle}. \nonumber
\end{equation}
Notice that the summation is restricted over the central core of the lattice $20 < i,j < 61$ where the occupation
of the upper level is appreciably large to avoid numerical problems by the division. $L_{\text{eff}} = 40$ is the effective size of this central region. This quantity
measures the time-evolution of the pair coherence uncluttered with pair density dynamics. 
We verified in the numerical simulation that a good accuracy of our results was obtained. The shown results were obtained with  a bond dimension up to $m=300$, a truncation error of $\epsilon_{trunc}=10^{-12}$ and a time step $Jdt = 0.0020\hbar$.

\end{document}